\documentstyle[12pt,aasms]{article}
\begin{document}
\begin{center}
\section*{\bf Non-thermal Origin of the EUV and Soft X-rays from the Coma
Cluster - Cosmic Rays in Equipartition with the Thermal Medium}
 
\vspace{5mm}
R. Lieu$^1$, W. -H. Ip$^2$, W. I. Axford$^2$, and M. Bonamente$^1$
 
\vspace{1.5mm}
 
$^1$ Department of Physics, University of Alabama, Huntsville, AL 35899. \\
$^2$ Max Planck Institut f\"ur Aeronomie, D-37191 Katlenburg-Lindau, Germany \\

\end{center}
 
\vspace{2mm}

{\it The role of cosmic rays (CR) in the formation and evolution of clusters 
of galaxies has been much debated.  It may well be related to other fundamental
questions, such as the mechanism which heats  and virializes the intracluster 
medium (ICM), and the frequency at which the ICM is shocked.  There is now 
compelling evidence both from the cluster soft excess (CSE) and the `hard-tail' 
emissions at energies above 10 keV, that many clusters are luminous sources of
inverse-Compton (IC) emission.  This is the first direct measurement of cluster 
CR: the technique is free from our uncertainties in the ICM magnetic field, and
is not limited to the small subset of clusters which exhibit radio halos.  The 
CSE emitting electrons fall within a crucial decade of energy where they have 
the least spectral evolution, and where most of the CR pressure resides.  
However their survival times do not date them back to the relic CR population.
By using the CSE data of the Coma cluster, we demonstrate that the CR are 
energetically as important as the thermal ICM: the two components are in 
pressure equiparition.  Thus, contrary to previous expectations, CR are a 
dominant component of the ICM, and their origin and effects should be explored.
The best-fit CR spectral index is in agreement with the Galactic value.}

\vspace{2.5mm}

Recent research on clusters of galaxies unveiled a number of independent
and contemporaneous indications that non-thermal activities 
in the ICM are at a much
higher level than previously thought.  First came the
EUVE discovery of 69 - 190 eV radiation in excess of that expected
from the thermal ICM,
confirmed
by the ROSAT and BeppoSAX detection of similar
soft X-ray (0.1 - 0.4 keV) excesses (Lieu et al 1996a,b;
Bowyer, Lampton and Lieu 1996; Fabian 1996; Mittaz, Lieu and Lockman 1998;
Bowyer, Lieu and Mittaz 1998; Kaastra 1998).  Details of the CSE data
suggest that the phenomeon is very plausibly due to  
IC emission by CR electrons scattering off the cosmic
microwave background (CMB; Ensslin and Biermann 1998; Hwang 1997; Sarazin and 
Lieu 1998), with the original thermal scenario being
eventually rejected because it requires an unrealistically large amount
of rapidly cooling gas (Mittaz, Lieu and Lockman 1998).
Since clusters are rarely
diffuse radio sources (Hanisch 1982)
the operation of IC is not
restricted in the most general approach (Sarazin and Lieu 1998) 
to the population of
radio synchrotron electrons only.
Secondly, the BeppoSAX discovery of hard X-ray tails in the spectra of 
several essentially randomly selected clusters (Kaastra 1998; Fusco-Femiano 
et al 1998) again
reveals that an active non-thermal ICM is commonplace.  In this {\it Letter}
we focus on the Coma cluster, which does possess a radio halo.  We
shall demonstrate for the first time that
the energetics of CR in this cluster are very important,
but the facts come from the latest data which are not related to
the radio properties of clusters.  
Our conclusion regarding the significance of CR is
therefore quite general.

We summarize how 
evidence of varying degrees of strength are now accumulating to form
a compelling case that Coma's CSE is indicative of intense
non-thermal activity: (a)
there is close resemblance in spatial morphology between
the EUV emission and the radio halo, which is not shared by the thermal
X-ray emission (Bowyer and Bergh\"ofer 1998); 
(b) the CSE data are modeled to the same satisfaction
with less parameters and components by a non-thermal spectrum (see below);
(c) extrapolation of the best-fit CSE power law model to hard X-ray energies
leads to a predicted flux which agrees very well with the BeppoSAX 
detected hard-tail flux
from this cluster (Fusco-Femiano et al 1998, again see below);
this greatly consolidates the IC
interpretation.

The EUV and soft X-ray data of Coma (Lieu et al 1996b) are modeled 
here with a power-law
plus thermal spectrum to respectively describe the CSE and the
X-ray emission.  The best parameters are shown in Table 1, where it can be
seen that reasonable goodness-of-fit is achieved using two {\it less} degrees
of freedom than pure thermal models with gas at warm and hot
temperatures (Lieu et al 1996b).  
A stringent test of the correctness of this model is
afforded by extrapolating the 
power-law to the 20 -- 80 keV
passband and comparing the total predicted flux with the cluster
hard X-ray excess flux measured by BeppoSAX (Fusco-Femiano et al 1998).  
The agreement is
within a factor of two.

The differential photon number index is
$\alpha \sim 1.75$ for all the annular regions, which transforms to a
similar index for the emitting electrons of $\mu \sim$ 2.5, consistent with
the index of $\mu =$ 2.7 for 
Galactic CR given the errors in $\alpha$ (Table 1).  
This, together with the fact that
the acceleration (viz., diffusive shock acceleration; Axford, Leer 
and Skadron 1977; Bell 1978a,b; Blandford and Ostriker 1978; Krymsky 1977)
and subsequent evolution (Ip and Axford 1985) of CR involve the same physics
at work in the ICM as those in the interstellar medium, strongly suggests
that when investigating non-thermal processes in clusters
our understanding of Galactic CR cannot be ignored.  In particular, limits
on the total CR pressure may be inferred from the CSE data.  Note that we
will not involve the hard X-ray and radio data, as
the pressure of these electrons
is much lower.  

Our ensuing conclusions regarding CR pressure
may be changed if the electrons do not
belong to the shock accelerated CR population within Coma's ICM.  
An alternative means of generating such highly relativistic
electrons is by the jets (especially e$^+$e$^-$ pair jets) 
emanating from Active Galactic
Nuclei (AGN).  This mechanism cannot account for the data because 
the long diffusion time of electrons in the ICM (V\"olk,
Aharonian and Breitschwerdt 1996) renders it difficult for AGN outflows to
form a large scale diffuse population of radiating electrons.  One
would require a widespread distribution of AGN, but the radio luminosity
function of clusters reveals too few energetic AGN to 
generate the CSE electrons.  The electrons will have to be relic ones,
injected at much earlier epochs when clusters had a larger number of AGN.
However, as demonstrated below, such electrons
would not have survived the severe interim losses.

In Figure 1 we show the time evolution of CR spectra in the environment
of the central region of Coma.  The initial spectra follow
those of Galactic CR at the time of production (which for our purpose may be
assumed instantaneous). 
In momentum space both
protons and electrons have a $p^{- \mu}$ power-law.
In energy space this transforms to $E^{- \mu}$ at relativistic energies
but flattens trans-relativistically, eventually to
to $E^{- \mu/2}$ at low energies.  Since the `flattening' occurs earlier
for protons, for equal number of both species
injected to the
shock at low energies the protons will outnumber the
electrons by a factor which is
$\gg$ 1 at energies
$\geq$ 0.1 GeV (a phenomenon quantitatively confirmed
by observations; Webber 1983)
and which increases with $\mu$ along with the
total pressure ratio of protons to electrons.  Note especially that for
an initial spectrum in 
the 0.1 - 1 GeV range relevant to the CSE, where the electrons are
relativistic and the protons are trans-relativistic, and the proton pressure
per decade of energy is maximized, this pressure is already higher than that
of the electrons by a factor of $\sim$ 20.

As time progresses the CR will suffer losses.  Escape is negligible for the
energy range of interest (V\"olk,
Aharonian and Breitschwerdt 1996) but $<$ 1 GeV particles will lose energy by
Coulomb collisions with the hot ICM 
(Sarazin and Lieu 1998).  More energetic protons are
preserved because their principal loss mechanisms: knock-on collisions,
neutron $\beta$-decay, and pion production, have interaction timescales (
Marscher and Brown 1978)
longer than a Hubble time.  However, energetic
electrons will be removed rapidly by IC and synchrotron losses (Sarazin and 
Lieu 1998),
resulting in abrupt spectral cut-offs (Figure 1).  The total pressure 
ratio, which
already has a large initial value, will therefore increase with time.

A parameter which places the present data in context is the ratio of the 
total proton pressure to the pressure of
CSE emitting electrons, with the latter defined as
electrons having Lorentz factors
between $\gamma_{min} =$ 300 and
$\gamma_{max} =$775, as they 
can undergo IC to produce photons within the CSE
energy range of 0.069 - 0.4 keV according to the formula $\gamma =
300 (h \nu_{CSE}/75 eV)^{1/2}$.
To begin with,
$r$ decreases slightly with time and reaches a broad minimum between
0.5 and 2 Gyrs (1 Gyr $=$ 10$^9$ years),
after that it rises sharply
(see Figure 2).  The minimum exists because after creation the total
proton pressure is rapidly reduced by the removal of low energy protons
whereas the CSE electrons, which straddle between the regime of Coulomb
and radiative losses, are in a band where the spectral evolution is minimized.
Nonetheless, for evolutionary timescales in excess of 2 Gyrs no significant
number of CSE electrons is expected, and the spectral slope is far steeper
than our observed value.
This places a {\it severe} limit on the age of the CR, and rules out
the possibility of the CR belonging to a relic population injected during
the supernova `bright phase' of z $\sim$ 2.

A potential effect on $r$ concerns
replenishment by secondary electrons as the pions decay.  Protons with energies
$\geq$ 1 GeV, which encompass $\sim$ 25 \% of the initial CR pressure
(Figure 1),
can interact with the hot ICM to produce pions; with the CSE emitting
electrons resulting from the subsequent pion decay carrying $\sim$ 1 \% of
the pressure of the $\geq$ 1 GeV protons (Marscher and Brown 1978).  
Thus $\sim$ 0.16 \% of
the CR pressure goes to creating CSE electrons within
the pion loss e-folding time of 150 Gyrs (Dennison 1980).  Given that synchrotron and
IC losses remove 
the initial $r \sim$ 1 \% electrons in $\sim$ 3 Gyrs,
evidently the pion decay process played a {\it negligible} role
in the evolution of the CSE electrons.
Similar calculations reveal that
replenishment is unimportant for electrons at all energies.
However, the decay of neutral pions
do result in a gamma ray flux $\sim$ 10 times below the current EGRET 
upper limit (Sreekumar et al 1996),
the detection of which would offer direct confirmation of the
existence of energetic protons at CR proportions.

An estimate of the total CR pressure in Coma is available from our
knowledge of $r$ developed
above, and from the pressure $P_e$ of the CSE electrons as determined
observationally via the equation $P_e^{CSE} = E_e^{CSE}/3V$, with
$V$ being the volume of a cluster region and $E_e^{CSE}$ related to
the deprojected CSE power-law luminosity $L_{CSE}^d$ by
\begin{equation}
E_e^{CSE} = 8 \times 10^{61} L_{42}^d \frac{3-\mu}{2-\mu} 
\frac{\gamma_{max}^{2-\mu} - \gamma_{min}^{2-\mu}}
{\gamma_{max}^{3-\mu} - \gamma_{min}^{3-\mu}}~~~\rm{ergs}
\end{equation}
where $L_{42}^d$ is
$L_{CSE}^d$ in units of 10$^{42}$ ergs s$^{-1}$.
The deprojection was performed by noting that the ratio
$L_{CSE}^p/L_X^p$ of observed (i.e. projected) luminosities does not
exhibit much variation with radius (see Table 1) and has
an average value of 0.127, implying that a similar
ratio of deprojected luminosities should also remain constant at this
value.  The ratio, coupled with the dependence of
$L_X^d$ on radius as obtained from the
$\beta$-model (Briel, Henry and B\"ohringer 1992), 
allows us to compute  $L_{CSE}^d$ and hence $P_e^{CSE}$.

The pressure ratio of thermal gas to
CSE electrons is shown in Table 1, with the former obtained directly from
the $\beta$-model.   The ratio varies from $\sim$ 200 at the cluster
center to $\sim$ 700 at an angular radius of 16.5 arcmin, with an overall
value of $\sim$ 250 for the entire sphere of radius 18 arcmin.
Now from Figure 2 the total CR pressure is determined from $P_e^{CSE}$ via
the value of $r$ at a given age of the CR, 
except this age is unknown because the epoch for the injection
of such a vast amount of CR remains to be explored (indeed this epoch
may not be 
unique).  Nonetheless, as a conservative estimate of the CR pressure, we
adopt the minimum value of $r$, which is
$\sim$ 150 for $\sim$ 1 Gyr of evolution,
Thus within 
the entire 18 arcmin radius
the overall CR to gas pressure ratio is $\geq$ 0.6, meaning that the two ICM
components are already in approximate equipartition.  

Can this profound consequence be avoided by varying certain aspects of the CR
model used here ?  The electron spectral index is consistent with that of
Galactic primary CR, and
is constrained by
the CSE data unless one assumes that the protons, which we do not measure, 
have a different index
from the emitting electrons.  However, theoretically the energetic protons
are expected to have a {\it smaller} value of $\mu$ since, unlike Galactic
ones, cluster protons cannot escape (V\"olk, Aharonian and Breitschwerdt 1996),
so if their spectrum
differs from Galactic it should actually
be closer to the $\mu =$ 2.2
index at the acceleration source (i.e. strong shocks).
Clearly a flatter proton index will increase the CR pressure further, so that
the current estimate is in fact a lower limit.


In conclusion, the IC interpretation of the CSE,
considered highly plausible because
of its theoretical self-consistency, the lack of conflict with earlier data
(especially measurements of cluster baryonic contents) and the ability to
explain the many varying behavior shown by
the recent data (soft excess and hard-tails), has 
significant implications on our understanding of clusters.  In particular,
in order to account for the observed CSE flux
of Coma, equipartition between CR
and gas is unavoidable.  This means the role of ICM CR (e.g. Berezinsky, 
Blasi, and
Ptuskin 1997) should be investigated in the light of the latest 
observations, as it has 
an impact on our understanding of cluster evolution; in particular
the gas heating process, and the history of the level of shock
acceleration activity 
necessary to produce such a vast amount of CR.  One question is whether 
mergers (Henry 1995) can shock the ICM
frequently and violently enough.
Further, more detailed modeling of an evolutionary
process which involves interactions between CR, gas, magnetic field
within the gravitational potential well of the dark matter is now
necessary.  Certainly the common expectation of the CR pressure being
$\sim$ 1 -- 10 \% of the gas pressure - an expectation based on the
past supernova rate as inferred from the iron abundance in the hot ICM - 
severely falls short of the current observations,
which represent the first direct measurement
of cluster CR.

The detection of non-thermal emission also causes a downward
revision of the gas mass, since the X-rays no longer have a
completely thermal origin.
For the entire 18 arcmin radius of Coma, 
our best-fit parameters indicate that the 
non-thermal component accounts for $\sim$ 22 \% of the
emission measure required by a purely thermal origin of the X-rays,
and its presence implies a reduction in the gas mass
by $\sim$ 11 \%.

Finally we note that our current efforts on the cluster Abell 2199 (paper in
preparation) involve simultaneous modeling of EUVE, ROSAT
and BeppoSAX data, and reveals a non-thermal component which rises in
prominence with cluster radius, consistent with the spatial behavior of
the CSE in this and two other regular clusters A1795 and A4038 (e.g. see
Mittaz, Lieu and Lockman 1998).  The presence of a dominant relativistic
electron population in the great voids of cluster fringes poses a major
puzzle, especially for the BeppoSAX hard-tail energies where the electrons
are extremely short-lived.
  
We thank Torsten Ensslin for helpful discussions.

\vspace{2mm}

\noindent
{\bf References}

\noindent
~Axford, W.I., Leer, E., Skadron, G. 1977, {\it Proc. 15th Int. Cosmic Ray
Conf.}, \\
\indent {\bf 11}, 132. \\
\noindent
~Bell, A.R. 1978a, {\it Mon. Not. Roy. Astr. Soc.}, {\bf 182}, 147--156. \\
\noindent
~Bell, A.R. 1978b, {\it Mon. Not. Roy. Astr. Soc.}, {\bf 182}, 443--455. \\
\noindent
~Berezinsky, V.S., Blasi, P., Ptuskin, V.S. 1997, 
\it Astrophys. J.\rm, {\bf 487}, 529--535. \\
\noindent
~Blandford, R.D., Ostriker, J.P. 1978, {\it Astron. Astrophys.}, {\bf 221},
L29. \\
\noindent
~Bowyer, S., Lampton, M., Lieu, R. 1996, {\it Science}, {\bf 274}, 1338--
1340. \\ 
\noindent
~Bowyer, S., Lieu, R., Mittaz, J.P.D. 1998,
{\it The Hot Universe: Proc. 188th \\
\indent IAU Symp., Dordrecht-Kluwer}, 52. \\
\noindent
~Bowyer, S., Bergh\"ofer, T. 1998, {\it Astrophys. J.}, in press
({\it astro-ph 9804310}). \\
\noindent
~Briel, U.G., Henry, J.P., B\"ohringer, H. 1992, {\it Astron. Astrophys.}, {\bf 259},\\
\indent L31--34. \\
\noindent
~Dennison, B. 1980, {\it Astrophys. J.}, {\bf 239}, L93. \\
\noindent
~Ensslin, T.A., Biermann, P.L. 1998, {\it Astron. Astrophys.}, {\bf 330},
90--98. \\
\noindent
~Fabian, A.C. 1996, {\it Science}, {\bf 271}, 1244--1245. \\
\noindent
~Fusco-Femiano, R., Dal Fiume, D., Feretti, L., Giovannini, G., Matt, G.,\\
\indent Molendi, S. 1998, {\it Proc. of the 32nd COSPAR Scientific Assembly, \\
\indent Nagoya, Japan (astro-ph 9808012)}. \\
\noindent
~Hanisch, R.J. 1982, {\it Astron. Astrophys.}, {\bf 111}, 97. \\
\noindent
~Henry, J.P., 1995, {\it Nature}, {\bf 377}, 13. \\
\noindent
~Hwang, C. -Y. 1997, {\it Science}, {\bf 278}, 1917--1919. \\
\noindent
~Ip, W. -H., Axford, W.I. 1985, {\it Astron. Astrophys.}, {\bf 149}, 7. \\
\noindent
~Kim, K. -T., Kronberg, P.P., Dewdney, P.E., Landecker, T.L. 1990, \\
\indent {\it Astrophys J.}, {\bf 355}, 29. \\
\noindent
~Krymsky, G.F. 1977, {\it Dok. Acad. Nauk. USSR}, {\bf 234}, 1306. \\
\noindent
~Lieu, R., Mittaz, J.P.D., Bowyer, S., Lockman, F.J.,
Hwang, C. -Y., Schmitt, \\ 
\indent  J.H.M.M. 1996a, \it Astrophys. J.\rm, {\bf 458}, L5--7. \\
\noindent
~Lieu, R., Mittaz, J.P.D., Bowyer, S., Breen, J.O.,
Lockman, F.J., \\
\indent Murphy, E.M. \& Hwang, C. -Y. 1996b, {\it Science}, {\bf 274},
1335--1338. \\
\noindent
~Marscher, A.P., Brown, R.L. 1978, {\it Astrophys. J.}, {\bf 221},
588--597. \\
\noindent
~Mittaz, J.P.D., Lieu, R., Lockman, F.J. 1998, {\it Astrophys. J.}, {\bf 
498},
L17--20. \\
\noindent
~Sarazin, C.L., Lieu, R. 1998, {\it Astrophys. J.}, {\bf 494}, L177--180. \\
\noindent
~Sreekumar, P. et al 1996, {\it ApJ}, {\bf 464}, 628. \\
\noindent
~V\"olk, H.J., Aharonian, F.A., Breitschwerdt, D. 1996, {\it Space Sci.
Rev.}, {\bf 75}, \\
\indent 279--297. \\
\noindent
~Webber, W.R. 1983, {\it Composition and Origin of Cosmic Rays, ed. M.M. \\
\indent Shapiro, Dordrecht-Reidel}, 83. \\


\vspace{2mm}

\noindent
{\bf Figure Captions}

\noindent
Figure 1: Time evolution of CR spectra in an ICM.  {\it 1a:} Proton
spectra at epochs 0, 0.5, 1.5, 3.0 and 4.5 Gyr from time of injection
(respectively curves 1 to 5).  {\it 1b:} As in {\it 1a} except for
electrons.   Note the y-axes of {\it 1a} and {\it 1b} have the {\it same}
arbitrary scale.
The Coulomb, synchrotron and inverse-Compton losses were
calculated according to the formulae in Sarazin and Lieu (1998), using a
gas density of 0.003 cm$^{-3}$ and a magnetic field strength of 2
$\mu$G.

\vspace{1mm}

\noindent
Figure 2: Time evolution of the pressure ratio of CSE emitting electrons
to all the CR protons.

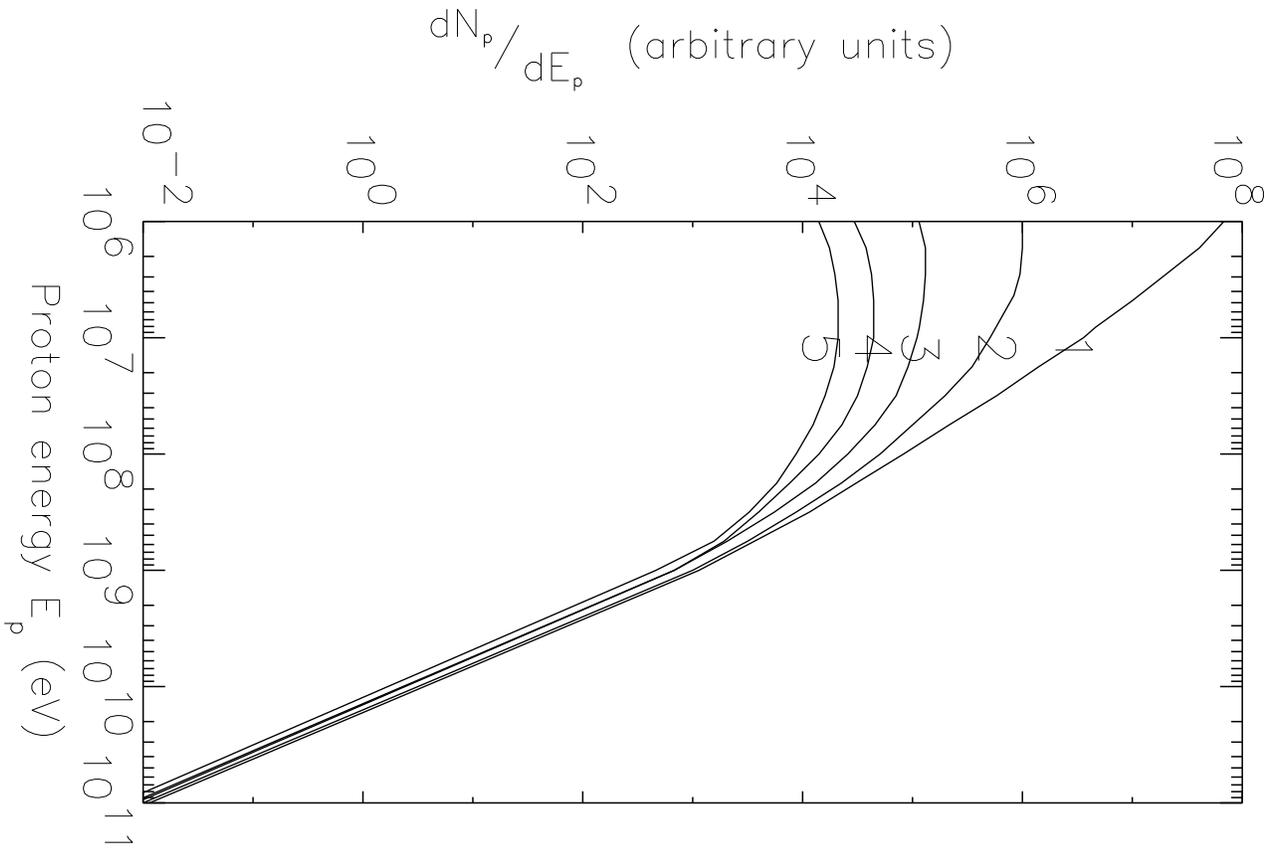
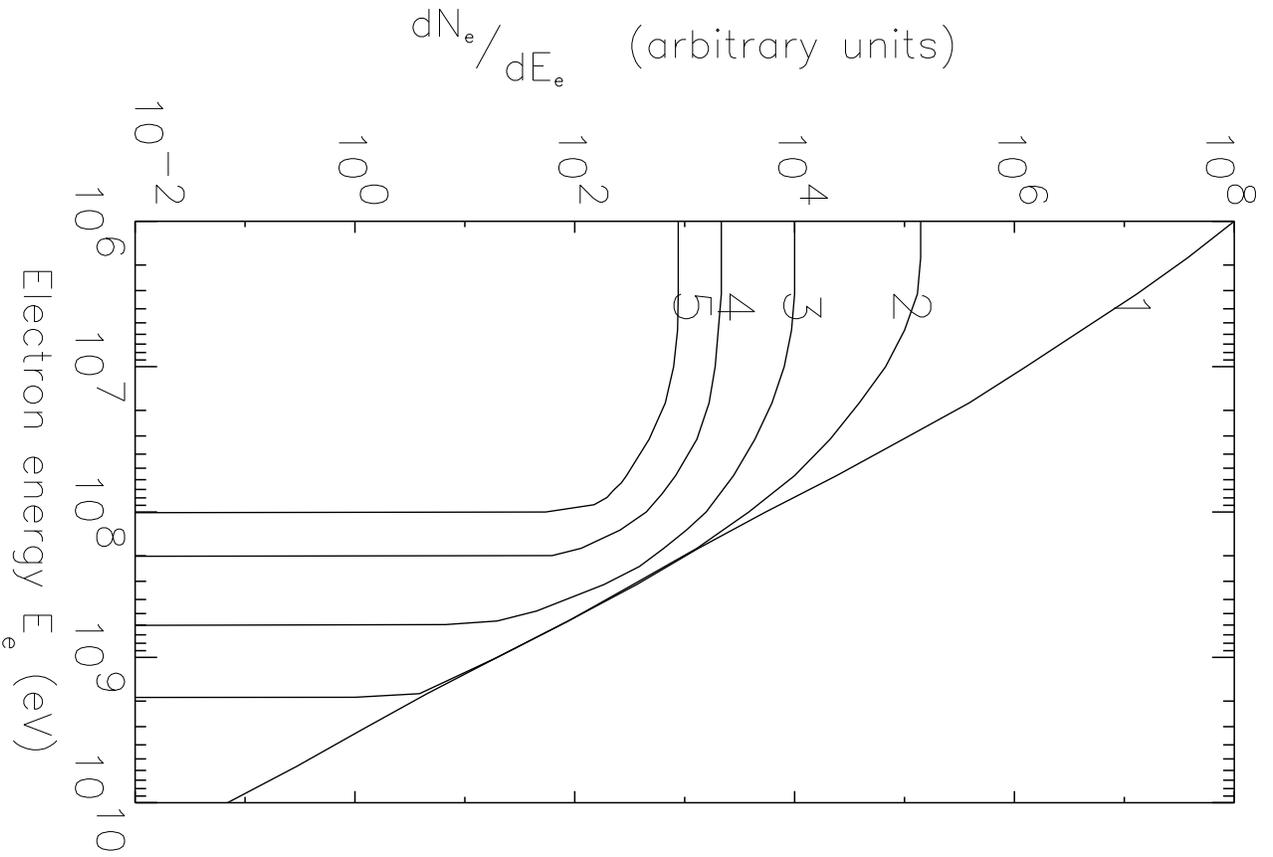
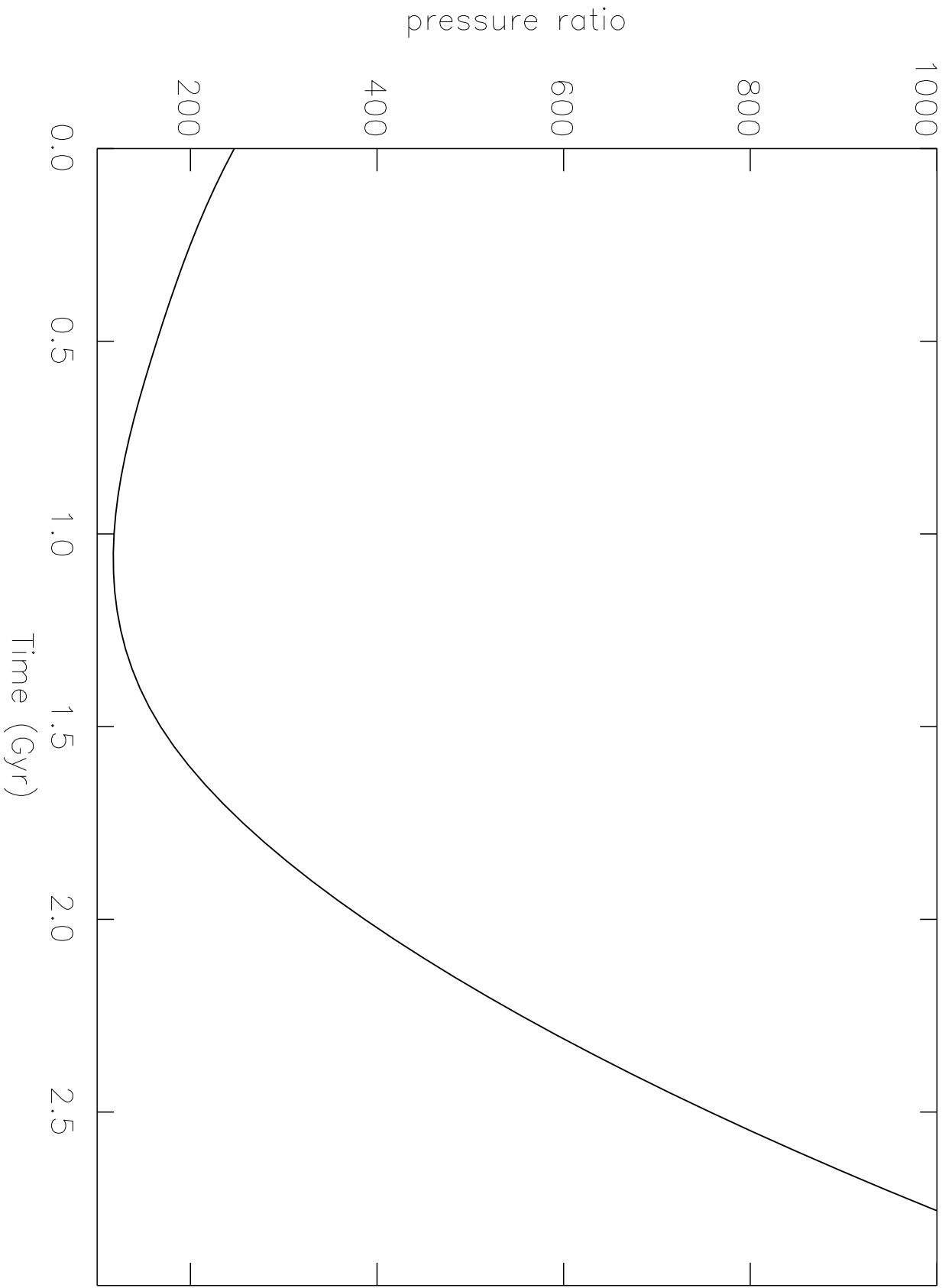
\end{document}